\documentclass[letter, 11pt]{revtex4}
\usepackage{epsfig,epstopdf,amsmath,amssymb,graphicx,subfigure,verbatim,natbib}

\usepackage{color}
\usepackage{calligra}
\DeclareMathAlphabet{\mathcalligra}{T1}{calligra}{m}{n}
\DeclareFontShape{T1}{calligra}{m}{n}{<->s*[2.2]callig15}{}

\newcommand{\ket}[1]{\left| #1 \right\rangle}
\newcommand{\bra}[1]{\left\langle #1 \right|}

\newcommand{\bv}[1]{\mathbf{#1}}

\begin{document}

\author{John A. Brehm}
\email{brehmj@sas.upenn.edu}
\affiliation{The Makineni Theoretical Laboratories, Department of Chemistry, University of Pennsylvania, 231 S. 34th Street, Philadelphia, Pennsylvania  19104-6323}

\author{Steve M. Young}
\email{smyoung@sas.upenn.edu}
\affiliation{The Makineni Theoretical Laboratories, Department of Chemistry, University of Pennsylvania, 231 S. 34th Street, Philadelphia, Pennsylvania 19104-6323}

\author{Fan Zheng}
\email{zhengfan@sas.upenn.edu}
\affiliation{The Makineni Theoretical Laboratories, Department of Chemistry, University of Pennsylvania, 231 S. 34th Street, Philadelphia, Pennsylvania 19104-6323}

\author{Andrew M. Rappe}
\email{rappe@sas.upenn.edu}
\affiliation{The Makineni Theoretical Laboratories, Department of Chemistry, University of Pennsylvania, 231 S. 34th Street, Philadelphia, Pennsylvania 19104-6323}

\date{\today}

\begin{abstract}
We calculate the shift current response, which has been identified as the dominant mechanism for the bulk photovoltaic effect,  for the polar compounds LiAsS$_\text{2}$,  LiAsSe$_\text{2}$, and NaAsSe$_\text{2}$.   We find that the magnitudes of the photovoltaic responses in the visible range for these compounds exceed the maximum response obtained for BiFeO$_\text{3}$ by 10 - 20 times.  We correlate the high shift current response with the existence of $p$ states at both the valence and conduction band edges, as well as the dispersion of these bands, while also showing that high polarization is not a requirement.   With low  experimental band gaps of less than 2 eV and high shift current response, these materials have potential for use as bulk photovoltaics.

\end{abstract}

\title{First-Principles Calculation of the Bulk Photovoltaic Effect in the Polar Compounds LiAsS$_\text{2}$,  LiAsSe$_\text{2}$, and NaAsSe$_\text{2}$} 
\maketitle

%\linenumbers

\section{Introduction.}

The bulk photovoltaic effect (BPVE) is the phenomenon in which electromagnetic radiation imparted on a single-phase insulating or semi-conducting material leads to a zero-voltage photo-current.  Like traditional photovoltaics,  ($e.$ $g.$  Si, CdTe, CIGS, and GaAs),  in order for a material to exhibit a significant BPVE response from  sunlight and thus be useful as a solar energy harvesting material, it needs to have a band gap in the visible spectrum (1.1 $\text{-}$ 3.1 eV) or the near-infrared.  Unlike traditional photovoltaics, which require an interface between two materials, the BPVE is achieved through the broken inversion symmetry in a single material.\cite{Young12p116601,vonBaltz81p5590,Sipe00p5337}   Additionally, only materials with nonzero polarization can give a current in response to unpolarized light, making them materials of interest for solar conversion.  This constraint stems from the physics of the non-linear optical process termed ``shift current," which we have demonstrated in our earlier theoretical works   is the dominant mechanism for generating the BPVE in the ferroelectrics BiFeO$_\text{3}$, BaTiO$_\text{3}$  and PbTiO$_\text{3}$\cite{Young12p236601, Young12p116601};  if a material is non-centrosymmetric but possesses no polarization,  then the directions of the generated shift currents from unpolarized light will sum to zero and produce no net current.\cite{Sturman92p1}  Many oxide perovskites have both of these properties, and the BPVE effect has been realized experimentally in them.\cite{Chen69p3389, Koch76p305, Choi09p63, Ji10p1763,Chakrabartty14pA80, Nechache11p202902,Grinberg13p509}  

Further,  based on the data from our studies on BaTiO$_\text{3}$ and PbTiO$_\text{3}$, we suggested that materials with elemental combinations conducive to covalent bonding and  delocalized electronic states can lead to large shift current effects.\cite{Young12p116601}   Our data also indicated that the magnitude of polarization is not simply proportional to the shift current produced.\cite{Young12p116601}  These observations have shaped our materials search.  We avoid  $A$$_\text{$a$}$$B_\text{$b$}$$X_\text{$x$}$ compounds with $B$-sites that have transition metals possessing localized conduction band electronic states, and concentrate on compounds with $B$-$X$ electronegativity differences less than one.  A natural set of $B$-$X$ combinations that meet these criteria are compounds with $B$-sites from Groups 14 and 15, and $X$-sites from Groups 16 and 17, except for O and F which have too high an electronegativity to meet the covalency requirement. In order to broaden the search, we remove the perovskite requirement of $a$ = $b$ = 1 = $x$/3.

\begin{figure}[h]
\vspace{0pt}
\includegraphics[scale = 0.15]{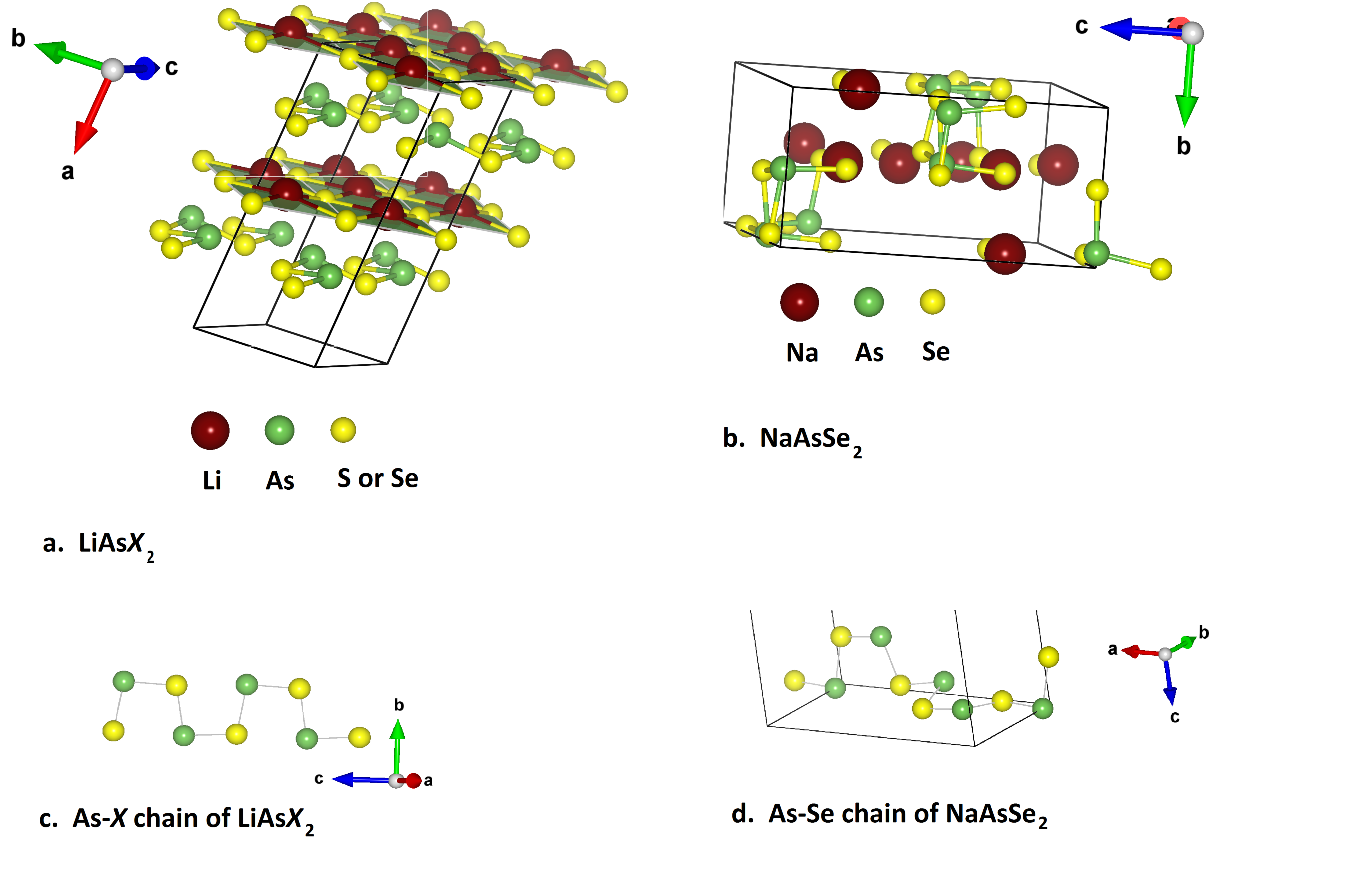}
\vspace{-20pt}
\caption{\label{one} Depictions of compounds  a) LiAsS$_2$ and LiAsSe$_2$, and b) NaAsSe$_\text{2}$.  c) As-$X$ chain in  LiAs$X$$_\text{2}$.  d) As-Se chain in NaAsSe$_\text{2}$.  The VESTA graphics software package was used to create these images.\cite{Vesta13p1} }
\end{figure}

In the current work, we calculate the BPVE of three ternary compounds that meet these criteria:  LiAs$X$$_\text{2}$ ($X$ = S, Se) and NaAsSe$_\text{2}$.  All three have been synthesized in polar monoclinic space groups:  $Cc$ for the first two  and $Pc$ for the third.\cite{Bera10p3484}  As well, all three compounds have been documented as having experimental band gaps well within the visible spectrum:  1.60 eV for LiAsS$_\text{2}$, 1.11 eV for LiAsSe$_\text{2}$, and 1.75 eV for NaAsSe$_\text{2}$.\cite{Bera10p3484}   These compounds are distinguished by their one dimensional infinite As-$X$  chains,  as shown in  Figure \ref{one}.  The chains in LiAs$X$$_\text{2}$ and NaAsSe$_\text{2}$ are different.  In  LiAs$X$$_\text{2}$, the chain atoms are confined to planes not containing Li, and the Li atoms arrange themselves  in a nearly square planar arrangement with the remaining non-chain $X$ atoms.  On the other hand, in NaAsSe$_\text{2}$, the Na atoms do not form square planar arrangements with Se.    The differences in the cation arrangements and the chain  are clearly visible in Figure \ref{one}.  Additional chain descriptions are detailed in Bera {\em et al.}\cite{Bera10p3484}  A final difference between the two types of compounds is that the $\beta$  angles, (between the $a$ and $c$ lattice vectors), in LiAsS$_\text{2}$ and LiAsSe$_\text{2}$ are 113.12$^\circ$ and 113.21$^\circ$, while  $\beta$ for NaAsSe$_\text{2}$  is  90.45$^\circ$, making this crystal  nearly orthorhombic.  In this paper, we report the calculated bulk photovoltaic shift current and Glass coefficient of these materials.

\section{Methodology.}

We use Quantum Espresso\cite{Giannozzi09p395502} to perform density functional theory calculations with the Perdew-Burke-Ernzerhof generalized gradient exchange-correlation functional on the three compounds described above.  We have found that calculations using experimental geometries, where available, allow for more faithful reproduction of electronic properties.  Additionally, the response obtained for relaxed structures tends to be stronger, so that the results shown are comparatively conservative.  We use the coordinates listed in the FIZ Karlsruhe ICSD database for LiAsS$_\text{2}$ and LiAsSe$_\text{2}$.\cite{Belsky02p364,FIZ13p1} The coordinates for NaAsSe$_\text{2}$ are taken from the supporting information of Bera {\em et al.}\cite{Bera10p3484}   The results of the SCF calculation are then used to calculate the partial density of states (PDOS) and band structure, and the wavefunctions and energies are also used as  inputs for the shift current calculation.  We use the nomenclature for the high symmetry points as found in the Bilbao Crystallographic Server to create band diagrams.\cite{Bilbao13p1}    We use ABINIT to calculate the polarization.\cite{Gonze02p478}  Norm-conserving optimized pseudopotentials\cite{Rappe90p1227} were created using the OPIUM software package.\cite{OpiumSourceforgenet11p1}  All calculations use a plane-wave basis set with a 50 Ry plane-wave cutoff.

As described in Refs. \cite{Young12p116601,vonBaltz81p5590,Sipe00p5337}, shift current is a second order, rectification-like optical process wherein charge is transported by coherent states that are created by interaction with light and allowed a net momentum as a consequence of inversion symmetry breaking. It may be derived using   time dependent perturbation theory under a dipole approximation treatment of the classical electromagnetic field. With $\bv{J}$ as the current density due to illumination with electric field strength $\bv{E}$, the response tensor $\sigma$ is expressed as:
\begin{flalign}
 J_q & = \sigma_{rsq}E_rE_s  \nonumber\\
 \sigma_{rsq}(\omega) & = \pi e \left(\frac{e}{m\hbar \omega}\right)^2 \sum_{n',n''} \int d\bv{k} \left(f[n''\bv{k}]-f[n'\bv{k}]\right)  \nonumber \\
                      & \quad \times \bra{n'\bv{k}}\hat{P}_r\ket{n''\bv{k}}\bra{n''\bv{k}}\hat{P}_s\ket{n'\bv{k}} \nonumber \\
                      & \quad \times \left(-\frac{\partial \phi_{n'n''}(\bv{k},\bv{k})}{\partial k_q}-\left[\chi_{n''q}(\bv{k})-\chi_{n'q}(\bv{k})\right]\right) \nonumber \\
                      & \quad \times \delta \left(\omega_{n''}(\bv{k})-\omega_{n'}(\bv{k})\pm \omega\right)
\end{flalign}

\noindent in which $n'$, $n''$, and $\bv{k}$ indicate band index and wavevector, $f$ gives the occupation, $\hbar$$\omega_n$ is the energy of state $n$, $\phi_{n',n''}$ is the phase of the momentum matrix element between state $n'$ and $n''$, and $\chi_n$ is the Berry connection for this state.

For the monoclinic space group compounds in this study, the shift current tensor is represented  in two-dimensional matrix form as:
    
\begin{eqnarray}
\sigma  = \begin{bmatrix}
       \sigma_{xxX} & \sigma_{yyX} & \sigma_{zzX} & 0 & \sigma_{xzX}  & 0 \\ 
       0 & 0 & 0 & \sigma_{yzY} & 0 &\sigma_{xyY}            \\[0.3em]
       \sigma_{xxZ} & \sigma_{yyZ} & \sigma_{zzZ} & 0 & \sigma_{xzZ}  & 0\\[0.3em]
     \end{bmatrix}
\end{eqnarray}

When the material is thick enough to absorb all the penetrating light, the Glass coefficient\cite{Glass74p233}  is used to describe the current response, and in the following we report only the terms diagonal in the field, from which the response to unpolarized light of an arbitrary wavevector may be determined. The absorption coefficient enters the Glass coefficient expression as $G_{rrq}=\sigma_{rrq}/\alpha_{rr}$, where $\alpha_{rr}$ is absorption coefficient tensor.   The shift current from a thick film can be expressed as:

%equation of glass coefficent
\begin{flalign}
J_q(\omega)  = \frac{\sigma_{rrq}(\omega)}{\alpha_{rr}(\omega)} \Big| E_r^0(\omega) \Big|^2 {\cal W}  = G_{rrq}(\omega)I_r(\omega){\cal W} 
\end{flalign}

\noindent where $I_r(\omega)$ is intensity and  $\cal{W}$ is the sample width.
Since we are, at present, concerned only with response to unpolarized light, we ignore terms off-diagonal in the electric field, as these cannot contribute to current.  To see this, we compute the general response in the $Z$ direction for unpolarized light with wavevector along $Y$. For arbitrary decomposition of the unpolarized light we obtain two orthogonal components
\begin{flalign*}
&\bv{E}'= E_0\left[\cos(\theta)\hat{\bv{x}}+\sin(\theta)\hat{\bv{z}}\right] \mbox{   and} \\
&\bv{E}''=E_0\left[-\sin(\theta)\hat{\bv{x}}+\cos(\theta)\hat{\bv{z}}\right]
\end{flalign*}
The current generated is then
\begin{flalign*}
J_z=&\left[\sigma_{xxZ}E'_xE'_x+\sigma_{zzZ}E'_zE'_z+2\sigma_{xzZ}E'_xE'_z\right] + E_0\left[\sigma_{xxZ}E''_xE''_x+\sigma_{zzZ}E''_zE''_z+2\sigma_{xzZ}E''_xE''_z\right]\\
J_z=&E_0^2\left[\sigma_{xxZ}\cos^2(\theta)+\sigma_{zzZ}\sin^2(\theta)+2\sigma_{xzZ}\cos(\theta)\sin(\theta)\right]+\\
       &E_0^2\left[\sigma_{xxZ}\sin^2(\theta)+\sigma_{zzZ}\cos^2(\theta)-2\sigma_{xzZ}\sin(\theta)\cos(\theta)\right]\\
J_z=&E_0^2\left[\sigma_{xxZ}+\sigma_{zzZ}\right]
\end{flalign*}
\noindent Thus, for unpolarized light, elements off-diagonal in the field will give canceling contributions.

\section{Results and discussion.}

\begin{table*}
\caption{\label{thedata} Calculated band gap, polarization, maximum shift current response, relative angle ($\gamma$) between the  $c$ lattice vector of the compound and $z$ polarization of incoming light at this maximum, and the maximum Glass coefficient at $\gamma$, and experimental band gaps  for LiAsS$_\text{2}$,  LiAsSe$_\text{2}$, and NaAsSe$_\text{2}$.  Values for BiFeO$_\text{3}$ are also reported. The experimental band gap values for the chalcogenide compounds are from Bera {\em et al.}\cite{Bera10p3484} }
\begin{ruledtabular}
\begin{tabular}{|c|c|cc|cc|ccc|} 
                 & Lattice&\multicolumn{2}{c|} {Band Gap} & \multicolumn{2}{c|} {Polarization} & Max. shift  & & Max. Glass \\ 
                 & $\beta$ angle&Calculated & Experiment & $P$$_\text{$x^\prime$}$ &$P$$_\text{$z^\prime$}$& current density&$\gamma$  & coefficient \\ 
Compound & ($^\circ$)&(eV) & (eV) & (C/m$^\text{2}$) & (C/m$^\text{2}$) & ($\times$10$^\text{-4}$ (A/m$^\text{2}$)/(W/m$^\text{2}$)) &($^\circ$) & ($\times$10$^\text{-9}$ cm/V) \\  \hline
NaAsSe$_\text{2}$	&\hspace{6pt}90.44&	1.25	&	1.75		&	-0.13	&	-0.06	&	109	& \hspace{5pt}0 & -35		\\
LiAsSe$_\text{2}$	&113.12&	0.77	&	1.11	&	-0.15	&	\hspace{4pt}0.06	&	-98 & 11	&	-42	\\
LiAsS$_\text{2}$	&113.25&1.07	&	1.60	&	-0.18	&	\hspace{4pt}0.06	&	-49	&	11 & -21	\\
BiFeO$_\text{3}$ & --- & \hspace{10pt}2.50\cite{Young12p236601} & \hspace{11.5pt} 2.67\cite{Basu08p091905} & 0 &\hspace{15pt} 0.90\cite{Wang03p1719}  & \hspace{20pt}5\cite{Young12p236601} & --- & \hspace{20pt}5\cite{Young12p236601}\\
\end{tabular}

\end{ruledtabular}
\end{table*}

\begin{figure}[h]
\vspace{30pt}
\includegraphics[scale = 0.35]{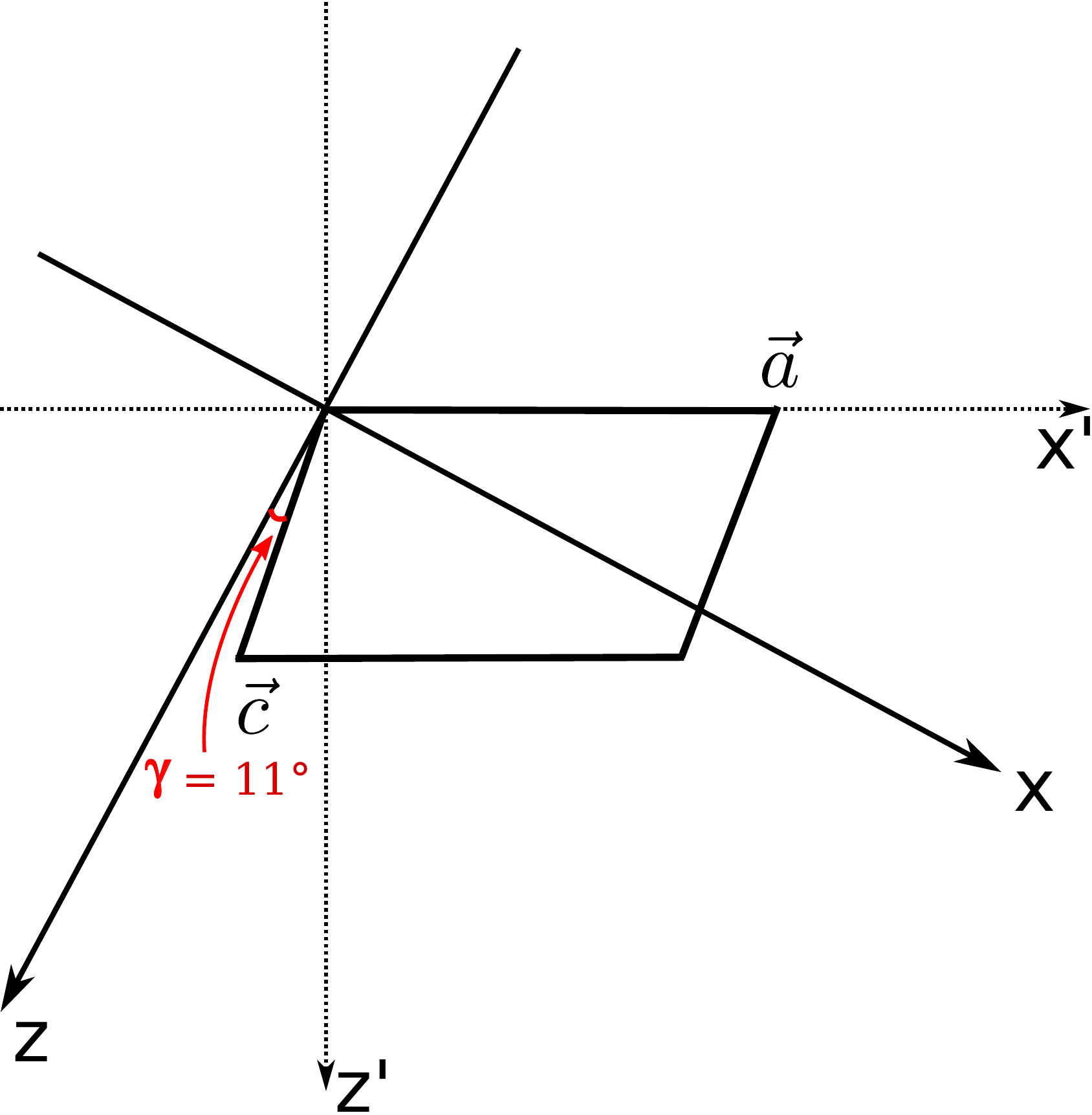}
\vspace{20pt}
\caption{\label{rot}Rotation of the LiAsS$_2$/LiAsSe$_2$ crystal in the x$^\prime$z$^\prime$ plane relative to incoming light for which the shift current response is a maximum.  The lattice vectors $\vec{a}$ and $\vec{c}$ are written in terms of $x^\prime$ and $z^\prime$, while the response and light polarizations are in terms of $x$ and $z$. The $zzZ$ response is maximized when the $z$ axis is rotated clockwise by $\gamma=11^\circ$ from $\vec{c}$.}
\end{figure}

\begin{figure}[h]
\vspace{0pt}
\includegraphics[scale = 1.50]{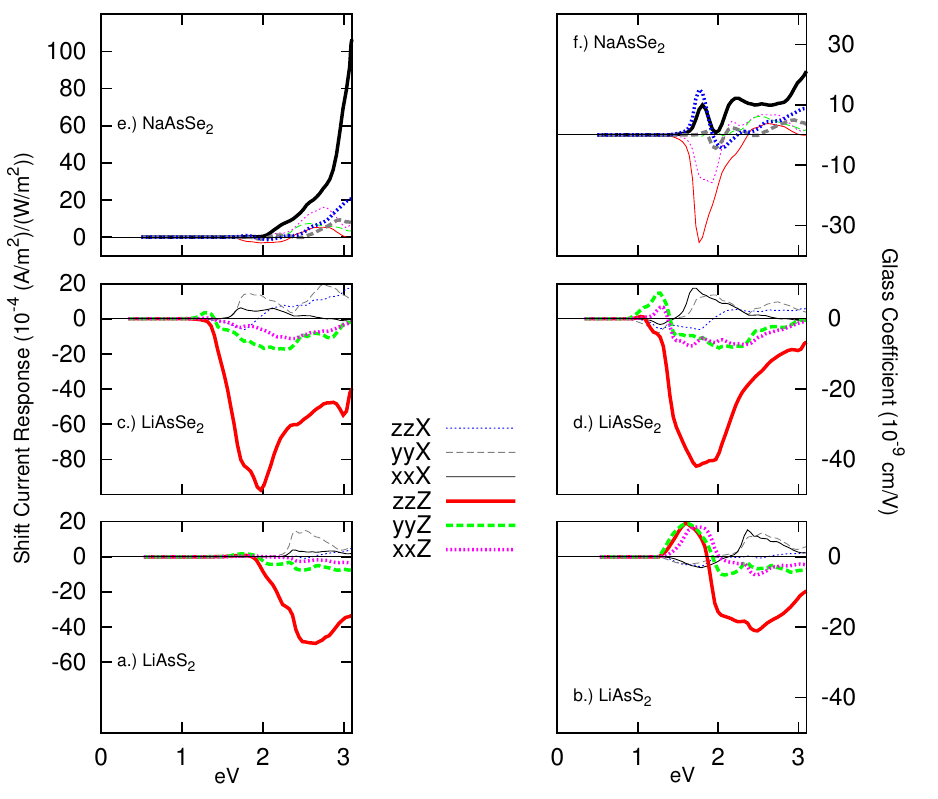}
\vspace{0pt}
\caption{\label{sss} Shift current responses and Glass coefficients for LiAsS$_\text{2}$,  LiAsSe$_\text{2}$, and NaAsSe$_\text{2}$.  The shift current responses are in the left hand column with units of $\times$10$^\text{-4}$ (A/m$^\text{2}$)/(W/m$^\text{2}$) and the Glass coefficient responses are in the right hand column with units of  $\times$10$^\text{-9}$ cm/V.   The response curves have been adjusted to the right by the difference in the experimental and calculated band gaps.  The legend entries are interpreted as follows:  $zzZ$ means polarized light from $zz$ direction inducing a current in the $Z$ Cartesian direction.}
\end{figure}

\begin{figure}[h]
\vspace{0pt}
\includegraphics[scale = 1.50]{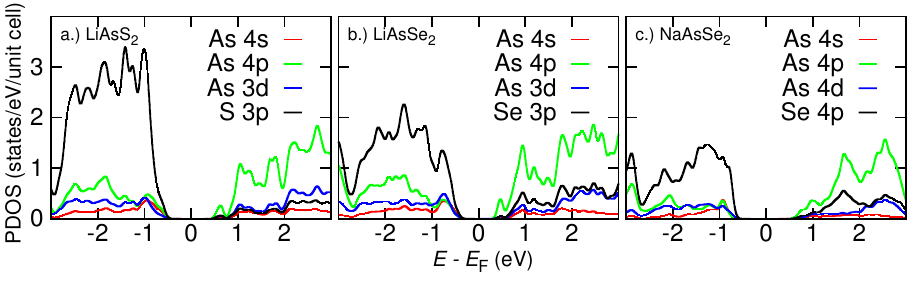}
\vspace{0pt}
\caption{\label{pdos1} PDOS for LiAsS$_\text{2}$,  LiAsSe$_\text{2}$, and NaAsSe$_\text{2}$.  For uniformity, the PDOS results are all relative to a 16-atom  unit cell.}
\end{figure}

\begin{figure}[h]
\vspace{0pt}
\includegraphics[scale = 1.50]{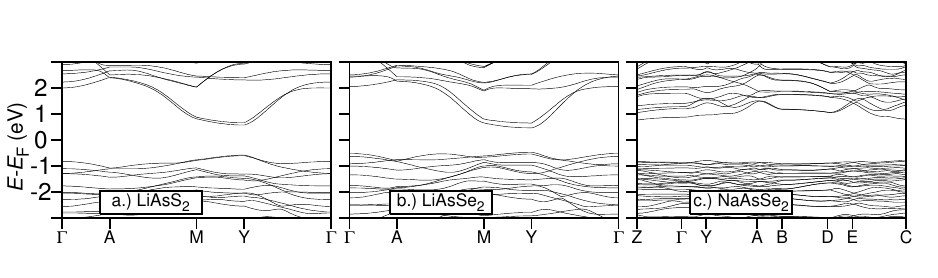}
\vspace{0pt}
\caption{\label{bands1} Electronic band structures for LiAsS$_\text{2}$,  LiAsSe$_\text{2}$, and NaAsSe$_\text{2}$.}
\end{figure}

Table \ref{thedata} presents the calculated maximum shift current density response and maximum Glass coefficient, as well as the calculated and experimental values for the band gap,  for the three compounds, ranked by maximum shift current response.   In order to show clearly the maximum responses, we rotate the lattices of the compounds counterclockwise in the $x^\prime$$z^\prime$-plane.  We define $\gamma$ to be  the angle between the $c$ lattice vector  and the $z$-component-aligned polarization of the incoming radiation at which maximum response occurs.    These values  are also listed in the table.  A cartoon of the orientation for LiAs$X$$_2$ ($X$ = S, Se) is provided in Figure \ref{rot}.  The nearly orthorhombic compound, NaAsSe$_2$, has its maximum shift current response at $\gamma$ = 0$^\circ$, while the $A$ = Li compounds, with nearly identical $\beta$ angles of 113.12$^\circ$ and 113.25$^\circ$, obtain maximum shift current response at $\gamma$ = 11$^\circ$. The bulk photovoltaic effect in BiFeO$_\text{3}$  has been the subject of significant recent study and interest, ($e.$ $g.$ Refs. \cite{Choi09p63,Ji10p1763,Alexe11p261,Bhatnagar13p2835,Nakashima14p09PA16,Yang10p143}), and thus serves as our benchmark.  It has been shown both theoretically and experimentally as having a maximum current density of 5$\times$10$^\text{-4}$ (A/m$^\text{2}$)/(W/m$^\text{2}$) at 3.3 eV and a maximum Glass coefficient of 5$\times$10$^\text{-9}$ cm/V at 2.75 eV.\cite{Young12p236601, Ji11p094115}  Each of the chalcogenide compounds in this study has at least an order of magnitude greater shift current response and Glass coefficient magnitude at least five times larger as well.

The total shift current responses and Glass coefficients are plotted in Figure \ref{sss} between 0 and 3.1 eV for $AB$$X$$_\text{2}$.  On each plot, the responses are shifted to the right by  the underestimation of the experimental band gap.  As depicted, the chalcogenides all show shift current responses for photon energies approximately 1 eV lower than the onset of the response for BiFeO$_\text{3}$, due to  their smaller band gaps.  At all energies below $\approx$2.9 eV, LiAsSe$_2$ has a superior shift current response and Glass coefficient to LiAsS$_2$ and NaAsSe$_2$. Above $\approx$2.9 eV, NaAsSe$_2$ has higher responses.     The responses are labeled such that the double small letters indicate the direction of the incoming radiation and the capital letter indicates the direction of the induced current.  (The reason for the two small letters is that the BPVE is a second-order process in the $E$ field.)   With respect to BiFeO$_\text{3}$ and its polarization value of 0.9 C/m$^\text{2}$,\cite{Wang03p1719} these data clearly reinforce our earlier work showing again that the magnitude of shift current is not simply  correlated with magnitude of material polarization.\cite{Young12p116601}

PDOS results in Figure \ref{pdos1} show that in each of the three compounds, the valence band edge down to -3 eV is dominated by S $3p$ or Se $4p$ states, while the  conduction band up to 3 eV is dominated by the As $4p$ states.  Thus, all electron transitions from the valence to the conduction band are overwhelmingly $p$ - $p$.  Band structures in Figure \ref{bands1} indicate that these three compounds all have direct band gaps.  The two compounds with $A$ = Li demonstrate significant dispersion in the conduction band.   Given the relative flatness of the conduction and valence bands in the vicinity of the band gap for NaAsSe$_\text{2}$, one would expect the other two compounds to have both smaller hole and electron effective masses, and hence higher mobility.   Thus, of the three compounds, we would expect LiAsSe$_\text{2}$ to produce the most current of the three compounds.

%This is unsurprising as As, Se, and S are remarkably similar in the extent of their repsective $p$ orbitals and the difference in energy levels between $4p$ As and Se is negliglible, and their difference versus $3p$ S is 1 Ry.   A correlation of response versus orbital energy then would suggest that minimizing the there is no one-to-one correspondence between the magnitude of the DOS at the valence and conduction band edges and the magnitude of the shift current response. 

\section{Conclusions.}
We have calculated maximum shift current responses in the visible range, adjusted for theoretical underestimation of experimental band gaps, in the range of 49-109$\times$10$^\text{-4}$ (A/m$^\text{2}$)/(W/m$^\text{2}$)   for LiAsS$_2$, LiAsSe$_2$ and NaAsSe$_\text{2}$.   The maximum shift current response values for LiAsS$_\text{2}$, LiAsSe$_\text{2}$ and NaAsSe$_2$  represent an order of magnitude improvement in response to visible light in comparison to BiFeO$_\text{3}$.  Glass coefficient responses are 4 - 8 times greater than that of BiFeO$_\text{3}$.  We note that expected efficiencies, while desirable, are not currently computable from first-principles.  To estimate efficiency, the photovoltage that can be sustained must be known; this depends strongly on the particular qualities of the sample and device as well as the bulk current density response, as observed in Refs. \cite{Alexe11p261,Bhatnagar13p2835}.  With band gaps below 2 eV, these non-perovskite, non-oxide compounds, with smaller polarization magnitudes than other oxide perovskites for which the BPVE has been evaluated, not only offer a higher shift current magnitude response, but capture more of the solar spectrum than BiFeO$_\text{3}$ as well.

\section{Acknowledgments}
JAB was supported by the DOE Office of Basic Energy Sciences, under grant DE-FG02-07ER46431.  SMY was supported by the AFOSR, under grant FA9550-10-1-0248.  FZ was supported by the NSF, under grant DMR-1124696.   AMR  was supported by  the Office of Naval Research, under grant ONR  N00014-12-1-1033.  Computational support was provided by the HPCMO of the U.S. DOD and NERSC Center of the U.S. DOE.

\end{document}